%% file: Infocomm2015_JournalVersion.tex
\documentclass[journal,10pt]{IEEEtran}
\usepackage{amsfonts}
\usepackage{mathrsfs}
\usepackage{amssymb}
\usepackage{wrapfig}
\usepackage{amsmath}
\usepackage{accents}
\usepackage{dsfont}
\usepackage{amsbsy}
\usepackage{setspace}
\usepackage{graphicx}
\usepackage{subfig}
\usepackage{url}
\usepackage{color}
\usepackage{epstopdf}
\usepackage{amssymb}
\usepackage{framed}
\PassOptionsToPackage{bookmarks={false}}{hyperref}

\newtheorem{theorem}{Theorem}

\newtheorem{algorithm}{Algorithm}

\newcommand{\mc}{\mathcal}

\begin{document}

\title{Low-Delay Distributed Source Coding for Time-Varying Sources with Unknown Statistics
\author{Fangzhou Chen, Bin Li and C. Emre Koksal}\thanks{This work is supported in part by NSF grant CNS-1054738.}\thanks{An Earlier version of this paper has appeared in the Proceedings of the 34th IEEE International Conference on Computer Communications (INFOCOM) Conference, Hong Kong, China, April, 2015 \cite{Fangzhou15}.}
\thanks{
Fangzhou Chen (chen.1953@osu.edu) and Can Emre Koksal (koksal@ece.osu.edu) are with the Department of Electrical and Computer Engineering at The Ohio State University, Columbus, Ohio 43210 USA. \newline 
\indent Bin Li (binli@uri.edu) is with the Department of Electrical, Computer and Biomedical Engineering at The University of Rhode Island,
Kingston, Rhode Island 02881 USA.}}

\maketitle

~\vspace{-0.3in} 

\input{Abstract}
\input{Intro}
\input{SystemModel}
\input{ProblemStatement}
\input{Approaches}
\input{Bounds}
\input{Experiment}

\input{Conclusion}

%

\input{Appendice}

\begin{spacing}{}
\bibliographystyle{abbrv}
\bibliographystyle{IEEEtran}
\bibliography{refs}
\end{spacing}

\end{document}

%% file: Abstract.tex
\begin{abstract}
We consider a system in which two nodes take correlated measurements of a random source with time-varying and unknown statistics. The observations of the source at the first node are to be losslessly replicated with a given probability of outage at the second node, which receives data from the first node over a constant-rate errorless channel. We develop a system and associated strategies for joint distributed source coding (encoding and decoding) and transmission control in order to achieve low end-to-end delay. Slepian-Wolf coding in its traditional form cannot be applied in our scenario, since the encoder requires the joint statistics of the observations and the associated decoding delay is very high. We analytically evaluate the performance of our strategies and show that the delay achieved by them are order optimal, as the conditional entropy of the source approaches to the channel rate. We also evaluate the performance of our algorithms based on real-world experiments using two cameras recording videos of a scene at different angles. Having realized our schemes, we demonstrated that, even with a very low-complexity quantizer, a compression ratio of approximately $50$\% is achievable for lossless replication at the decoder, at an average delay of a few seconds.
\end{abstract}
\begin{IEEEkeywords}
Lossless distributed source coding, universal algorithms, delay optimal control, heavy-traffic analysis
\end{IEEEkeywords}

%% file: Intro.tex
\section{Introduction}\label{Intro}
In many applications, multiple nodes take measurements from the same source to be combined later in order to obtain a high resolution representation of the source. In order to achieve that, nodes encode their digitized observations and share information for the observations to be replicated at a common location. Lossless distributed source coding aims to encode the observations in a way to minimize the rate of exchanged information for all observations to be perfectly replicated at a certain location. The nodes exploit the correlations in their observations to build an efficient code. Following the seminal work by Slepian and Wolf~\cite{SW}, there has been a vast interest in lossless distributed source coding (DSC) including applications in quantum key distribution~\cite{RM}, distributed video coding~\cite{Girod} and wireless sensor networks~\cite{Xiong}.

We address the lossless distributed source coding problem for a pair of nodes, observing a random source with time-varying statistics, unknown to the nodes before the session starts. Our objective is to minimize the delay for the second node to losslessly replicate the observations of the first node, subject to a given desired probability of outage. Communication from the first node to the second node occurs over a finite-rate channel. Slepian-Wolf (SW) coding in its traditional form cannot be applied in our scenario, since the encoder requires the joint statistics of the observations causally to encode information. The alternative is to encode across a time period, long enough to exploit long-term variations of the source. However, with this approach, the corresponding decoding delay is very high. To that end, we develop a system and associated novel strategies for joint distributed source coding (encoding and decoding) and transmission control in order to achieve low end-to-end delay. We evaluate the performance of our strategies both analytically and via real-world experimentation. We first derive upper and lower bounds on the expected delay. Our bounds show that the delay achieved by our strategies are order optimal as the conditional entropy of the source (given the observation at the second node) approaches to the channel capacity. Next, we use two cameras recording videos of a common scene at different angles to obtain experimental data. After evaluating the possible joint source distributions based on the observations, we apply our schemes and show major improvements in end-to-end delay over existing traditional Slepian-Wolf based coding schemes. The {\em analytical} results show us that the delay performance of our schemes are very close to that of a highly-optimistic imaginary scenario, in which the joint distribution of the source observations are causally available to encoder, without seeing the observation at the decoder. The {\em experimental} results show that, even if we use highly coarse quantization for the source statistics, the average data rate at which an encoder shares information to achieve lossless replication at the decoder is reduced by $\sim 50$\% (compression ratio), at an average end-to-end delay of $6$-$9$ secs.

\begin{figure}[t]
     \center
     {\includegraphics[scale=0.3]{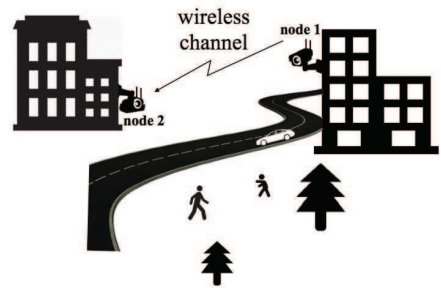}}
     \vspace{-0.2in}
    \caption{A sample scenario for our problem.}
\vspace{-0.425in}
     \label{EF}
\end{figure}


\noindent {\bf Sample scenario:} In Fig.~\ref{EF}, we provide an example for a typical setting\footnote{In fact, the illustrated scenario is the very case we evaluate in our experimental observations.} that we consider in this paper. Here, two security surveillance cameras observe the action in the environment, which is time-varying and uncertain. Therefore, the joint distribution of the observations of the two cameras is not causally available at these two locations. One of the nodes, holding a camera would like to replicate the video taken by the other camera to increase the resolution, in order to enhance the detection performance of an intruder, for example. The brute force approach would be to have the whole video transmitted to the replicator node over the wireless channel. However, this would lead to a waste of communication resources, especially if the wireless channel has a limited capacity. Instead, the first node can exploit the existence of highly correlated observation at the second (replicating) node to send information merely ``sufficient'' for its video to be replicated. This process is known as lossless distributed source coding. In security applications, delay is of critical importance. Therefore, traditional schemes that are based on the utilization of the long-term statistical regularities are not viable. Furthermore, due to time-varying statistics (non-stationarity), the knowledge of the joint statistics at the encoder is also not a valid assumption, which calls for {\em universal} solutions for coding.

\noindent {\bf Background, Related Work, and our Contributions:} Lossless distributed source coding was first introduced by Slepian and Wolf in~\cite{SW}. There, two independently and identically distributed (i.i.d.) random source sequences, $\{X_i\}_{i=1}^{\infty}$,$\{Y_i\}_{i=1}^{\infty}$  with joint cumulative distribution function (cdf) $F_{XY}(x,y)$ are observed at the encoder and the decoder, respectively. The objective is to reconstruct sequence $\{X_i\}_{i=1}^{\infty}$ losslessly at the decoder. Instead of encoding $\{X_i\}_{i=1}^{\infty}$ with its full entropy rate $H(X)$, SW coding enables lossless replication at an encoding rate $R_X=H(X|Y)$, regarding $\{Y_i\}_{i=1}^{\infty}$ as side information at the decoder. To achieve this rate, SW coding (i) encodes the source sequence across infinitely many blocks to achieve an arbitrarily low probability of decoding error; (ii) requires the knowledge of the joint cdf, $F_{XY}$ at the encoder. These two assumptions make direct application of SW coding, but are inappropriate in some delay-sensitive applications (e.g. live video meeting, online video streaming etc.) or for cases in which the joint statistics is unknown, and/or time-varying/non-stationary in certain situations (e.g. the encoder and decoder are moving while communicating).


These issues were addressed in a variety of studies, following~\cite{SW}.
Many practical coding schemes have been proposed, e.g. \cite{Puri02,Stankovic06}. And most of them are based on channel codes \cite{Stankovic04}, especially low-density parity check (LDPC) codes \cite{Liveris02,Matsuta10}. Nevertheless, perfect knowledge of joint cdf remains a widely adopted assumption. Csiszar and Korner first extended SW coding to achieve universality~\cite{CK}, where neither encoding nor decoding depends on source statistics. They also analyzed the finite-block behavior and provided the universally attainable error exponent when encoding rates are within the SW-region, i.e., the region of rates for which lossless replication of sources is possible with arbitrarily low probability of outage as the block size goes to infinity. In~\cite{Cheng}, end-to-end delay is studied in a SW coding setup. There, decoding error exponent is derived for a given end-to-end delay. If a feedback channel from the decoder to the encoder is present,~\cite{Reza} proposed a scheme under lack of the knowledge for the joint cdf. There, by carefully choosing a sequence of codes, the average encoding rate is minimized over each frame at the encoder. Similarly,~\cite{Jaggi} exploited the feedback channel to send some information to the encoder for a better performance. With unknown statistics,~\cite{SS} studied the outage capacity of the system and~\cite{Elsa} proposed a compound source model to achieve arbitrarily low probability of outage. In another direction,~\cite{Draper} proposed universal incremental SW coding, in which an incremental transmission and a universal sequential decision test is applied by the encoder and decoder, respectively. The system developed requires the availability of ACK/NAK feedback from decoder to encoder.

%

None of the above studies addressed the time varying and unknown source statistics in the context of delay minimization, without an active use of a feedback channel. Indeed, if the prior knowledge of the source statistics is not available, it may not be possible to feed back the joint statistics without actually feeding back the actual observation sequence itself. With that motivation, the main contributions of our paper can be listed as follows:

\noindent {\bf (1)} We extend the existing studies on universal distributed source coding by integrating time-varying and unknown joint statistics, the use of finite-capacity channel from the encoder to the decoder, and minimization of end-to-end delay.

\noindent {\bf (2)} We specify two different classes of joint encoding, decoding, and transmission control strategies, named Wait-to-Encode and Wait-to-Decode, and develop a strategy in each class to achieve low-delay in universal distributed source coding.

\noindent {\bf (3)} We derive upper and lower bounds on the performance of our strategies. While the bounds are valid in any regime, in the heavy traffic limit as the channel capacity converges to the long-term average source rate, we show that our strategies achieve optimal delay scaling. To achieve that, we develop and utilize new techniques in heavy-traffic analysis of queues. We also point out a phase-transition phenomenon on the achievable delay scaling to show that even a minor degradation in the knowledge of joint statistics leads to a different scaling regime.

\noindent {\bf (4)} We implement our stategies in a setup that involves two cameras recording videos of a common scene at different angles. This setup enabled the real-world experimental evaluation of the performance of our strategies. We demonstrated that, even with a low-complexity quantizer, a compression ratio of $\sim 50$\% is achievable for lossless replication at the decoder, at an average delay of a few seconds.

%% file: SystemModel.tex
\section{System Model}\label{SysM}
In this paper, we use boldface to represent vectors, upper case to represent random variables and vectors, and lower case to represent realizations of random variables and vectors, or deterministic parameters.

First, we introduce our {\bf source model}. In our system, there are two nodes that take correlated measurements of a random source. Time is slotted and a single source symbol is taken by each node in each time slot. We further group time slots into blocks of size $n$ time slots and denote the random measurement taken by the nodes in block $t$ as $\mathbf{X}_t$ and $\mathbf{Y}_t$, respectively. Hence, we will refer to the nodes as node $X$ and node $Y$ in the sequel. Source symbols are discrete, taking on values from associated finite sets $\cal{X}$ and $\cal{Y}$, for node $X$ and node $Y$, respectively. We denote the joint cumulative distribution function (cdf) of the observed source symbols in block $t$ with $F_{\mathbf{X}_t\mathbf{Y}_t}(\mathbf{x},\mathbf{y})$, and the associated probability mass function (PMF) with $P_{\mathbf{X}_t\mathbf{Y}_t}(\mathbf{x},\mathbf{y})$. We assume the joint statistics of the blocks to be time varying from one block to another, but to remain constant within each block (analogous to block fading models in channel coding) of $n$ symbols, large enough to invoke random source coding arguments. Also, the symbols observed by node $X$ and $Y$ are i.i.d. in each block: 
\begin{align*}
F_{\mathbf{X}_t\mathbf{Y}_t}(\mathbf{x},\mathbf{y})&=\prod_{i=1}^{n} F_{X_tY_t}(x_i,y_i), \\
P_{\mathbf{X}_t\mathbf{Y}_t}(\mathbf{x},\mathbf{y})&=\prod_{i=1}^{n} P_{X_tY_t}(x_i,y_i),
\end{align*}
where $F_{X_tY_t}(x,y)$ and $P_{X_tY_t}(x,y)$ denote the joint cdf and the associated PMF of a symbol pair in block $t$, respectively. In the sequel, we simply use $F_{(t)}(x,y)$ and $P_{(t)}(x,y)$. We further assume that $F_{X_tY_t}(x,y)$ takes values from a finite set $\cal{F}$, of possible joint cdfs. Without loss of generality, we group the joint cdfs in $\cal{F}$ into $m$ groups as follows: $\cal{F}$ $ = \{ F_{11},\dots F_{1l_1}; F_{21},\dots F_{2l_2}; \dots; F_{m1},\dots F_{ml_m} \}$, where each group, $\{F_{ij}\}_{j=1}^{l_i}$, of joint cdfs have identical marginal cdfs, i.e., $F_i(x)\triangleq F_{i1}(x,\infty)=\cdots=F_{il_i}(x,\infty),F_i(y)\triangleq F_{i1}(\infty,y)=\cdots=F_{il_i}(\infty,y)$ for all $i, 1 \leq i \leq m$. Since nodes $X$ and $Y$ merely have their own observations $\mathbf{X}_t$ and $\mathbf{Y}_t$, they only have the knowledge of the marginals $F_{X_t}(x)$ and $F_{Y_t}(y)$, available at the end of block $t$, but they do not have any knowledge of $F_{X_tY_t}(x,y)$ beyond the possible group that it belongs to $\mc{F}$. However, the nodes have the knowledge of the PMF, $P_F(F_{ij})$ of each possible joint cdf of the node. Note that, the existence of such a PMF implies a doubly stochastic, stationary, and ergodic structure in our source model: the joint cdf in each block $t$ is chosen at random, i.i.d. with probabilities parameterized by PMF $P_F(F_{ij})$. Then the source symbols are chosen at random, according to the associated joint cdf $F_{ij}$.

\begin{figure}
\vspace{-0.0in}
\begin{center}
{\includegraphics[height=0.85in]{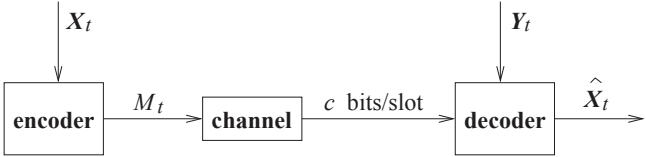}} \vspace{-0.0in}
\caption{System model.}
\label{fig:dnet}
\end{center}
\vspace{-0.35in}
\end{figure}

Next, we present the {\bf system} that we consider, as illustrated in Fig.~\ref{fig:dnet}. Node $X$ is connected to node $Y$ via a noiseless channel with constant transmission rate $c$ bits per slot. Node $X$ (source) encodes the observed symbols $\{\mathbf{X}_t\}$ into a string of bits, which we refer to as the message. We denote the message created in block $t$ with $M_t$. For each block, node $X$ and $Y$ also (possibly) exchange the indices of the marginal distributions (between $1$ and $m$)\footnote{This consumes a negligible amount of resources, compared to the size of the message, since $n$ is assumed to be large.}, i.e., node $X$ sends the index of $\mathbf{X}_t$ and node $Y$ sends the index of $\mathbf{Y}_t$.  Note that, to form a message, the encoder can possibly combine multiple blocks of source symbols. Consequently, there exist times in which the encoder chooses to make further observations to combine with the existing ones and not encode the current block at the time. For such blocks, a blank message is generated. For instance, in traditional SW coding, the encoder waits to observe infinitely many blocks of symbols that are encoded jointly to take advantage of long-term statistical averaging. The size of the string dictates the instantaneous rate of the encoder, denoted by $R_t$ bits/slot. Thus, the total number of bits in message $M_t=nR_t$. In case of a blank message after block $t$, $R_t=0$ for that block. We ignore the number of bits the encoder uses to encode the index of $\mathbf{X}_t$ in calculating the rate (i.e., $\log_2 m \ll nR_t$ for all non-blank messages). These messages are transmitted over the channel to be (source) decoded by the decoder at node $Y$. The decoder combines the received messages with its own observation sequence $\{\mathbf{Y}_t\}$ (as well as the indices of the marginals of blocks $\mathbf{X}_t$ if available) in order to {\em losslessly} decode $\{\mathbf{X}_t\}$. We denote the decoded sequence with $\{\mathbf{\hat{X}}_t\}$. If for certain block $t$, $\mathbf{\hat{X}}_t \neq \mathbf{X}_t$, we say that an outage occurred for block $t$.

We refer to a strategy as a method that jointly selects the encoder and the decoder. In particular, a strategy, parameterized with $\pi$, chooses the mapping from the sequence of blocks $\{\mathbf{X}_t\}$ observed thus far to the message $M_t$ , and the mapping from the received sequence $\{M_t\}$ to the decoded blocks $\{\mathbf{\hat{X}}_t\}$, at the end of each block. The set of all strategies is denoted with $\Pi$. Moreover, we do not impose any restriction on the strategy space such as stationarity or ergodicity.

Finally, we provide the {\bf delay model}. We measure the delay experienced by a block of source symbols as the time elapsed between the slot that the first source symbol is observed and the slot that all the symbols of the block is decoded. There are three different components of the delay that experienced by $\mathbf{X}_t$. Firstly, for a given strategy $\pi$, a block may experience a delay, $W_E^{(\pi)}(t)$, at the encoder. This is due to the fact that the encoder decides to group the symbols of the block with the symbols of the subsequent blocks. Next, the messages (encoded symbols) need to wait to be transmitted over the channel, since the channel has a finite rate. For instance, if the message at time $t$ has a rate $R_t=2$ Mbits/slot and the channel has a rate $c=1$ Mbits/slot, then it takes the message (encapsulating all blocks encoded) at least $2n$ slots (or $2$ blocks) to be transmitted over the channel, even if there is no other message in transmission at the transmission queue when the generated message $M_t$ arrives. We denote the transmission delay associated with a strategy $\pi$ with $W_C^{(\pi)}(t)$. Lastly, depending on the encoder strategy, the decoder may choose to accumulate further information on a source block through future messages and thus defer the decoding decision until later. We denote the decoding delay associated with strategy $\pi$ with $W_D^{(\pi)}(t)$. The overall delay experienced in the system with strategy $\pi$ is thus, $W^{(\pi)}(t)=W_E^{(\pi)}(t)+W_C^{(\pi)}(t)+W_D^{(\pi)}(t)$.

We finalize the section noting that, systems studied in \cite{Reza} and \cite{Cheng} can be regarded as two special cases of our system.

%% file: ProblemStatement.tex
\section{Problem Statement}\label{PS}
In this paper, our objective is to develop strategies that minimize the end-to-end delay observed by the source, while keeping the rate of blocks experiencing outage below a certain desired threshold $\epsilon\in(0,1)$. This goal can be achieved by solving the following problem:
\begin{eqnarray}
\min_{\pi \in \Pi}   &\limsup \limits_{T \to \infty} \frac{1}{T} \sum\limits_{t=1}^{T}\mathbb{E}[{W_E^{(\pi)}(t)+W_C^{(\pi)}(t)+W_D^{(\pi)}(t)}] \label{Opt1} \\
\text{s.t.}  \nonumber      &{\mathbb P} \{\mathbf{\hat{X}}_t \neq \mathbf{X}_t\} \leq \epsilon, \forall t=1,2,..., \infty,
\end{eqnarray}
where the expectation is taken over the PMF, $P_F(F_{ij})$, of the source cdfs and the probability of outage is dictated by the strategy $\pi$, as well as $P_F(F_{ij})$. Note that, while the source is stationary and ergodic, the strategies need not be stationary nor ergodic. One thing to be careful about our formulation is that, the outage probability is imposed on {\em every single} block individually, rather than on average.

\noindent {\bf Illustrative scenarios:}

\noindent (1) {\em Known joint cdfs:} It is well-known that if the joint cdf, $F_{(t)}(x,y)$, of the source were known at the encoder, then one could apply SW encoding and decoding~\cite{SW} (based on random binning and typicality decoding) on a block-by-block basis. Thus, the rate of the message $M_t$ would be the conditional entropy, $H_{(t)}(X|Y)$, of the source associated with joint cdf $F_{(t)}(x,y)$, which would lead to a long-term average encoding rate of
\[ \mathbb{E}[H_{(t)}(X|Y)]=\sum_{1 \leq i \leq m}\sum_{1 \leq j \leq l_i}{P_F(F_{ij})\cdot H_{ij}(X|Y)},\]
where $H_{ij}(X|Y)$ is the conditional entropy given $F_{ij}$.

With this approach, an arbitrarily low probability of outage can be achieved as $n \to \infty$ and thus the constraint is met. The encoding delay is merely a single block for all $t$, since each block is immediately encoded. The decoding delay is $0$ for all $t$, since each block is immediately decoded. Thus, the only component of the delay experienced is the transmission delay, which is finite if $c>\mathbb{E}[H_{(t)}(X|Y)]$.

\noindent (2) {\em Accumulate and encode:} Without the knowledge of the joint cdfs, one possibility is to accumulate infinitely many blocks at the encoder and encode them jointly. That way, node $X$ can exploit the law of large numbers as the empirical PMF of the source statistic, converges to $P_F(F_{ij})$ with probability 1. Thus, the situation becomes that of known joint statistics, and the long-term average encoding rate of $\mathbb{E}[H_{(t)}(X|Y)]$ can still be achieved at an arbitrarily low decoding error probability. However, clearly the encoding delay will be arbitrarily large with this approach and it cannot be a viable solution for our problem.

\noindent (3) {\em Block-by-block encoding:} In the other extreme, where each block is encoded separately, one has to pick the encoding rate large enough to ensure the outage constraint is met in each block. For instance, to achieve arbitrarily low probability of outage, the encoding rate has to be picked as
\[ R_t=\max_{1\leq i\leq m,1\leq j\leq l_i}{H_{ij}(X|Y)}\triangleq H_{\max}(X|Y).\]
While the encoding delay is a single block, the highly conservative choice of encoding rate increases transmission delay significantly, potentially to $\infty$ if $c < H_{\max}(X|Y)$.

The above observations motivate us to find a solution somewhere in between the two extreme approaches depicted in (2) and (3). Note that, if $c \gg \mathbb{E}[H_{(t)}(X|Y)]$ the problem becomes uninteresting (trivial if $c \geq H_{\max}(X|Y)$). The problem becomes interesting for $\mathbb{E}[H_{(t)}(X|Y)]<c<H_{\max}(X|Y)$. The major focus in performance analysis will be the case in which $c$ is very close to, but slightly larger than $\mathbb{E}[H_{(t)}(X|Y)]$.

While the extension to the case of the time-varying/fading channels is possible, the nature of the proofs change substantially and the derivations become cumbersome, leading to a loss in the main insights. Besides, our main objective in this paper is to address the delay caused by the variations in the source, as opposed to the temporal variations in the channel. Therefore, we chose to use static channels in the sequel.

%% file: Approaches.tex
\section{Proposed Approaches}\label{Approaches}
With the mere assumption that ${\cal F}$ is a finite set, there does not exist a well-structured closed-form solution for optimization problem (\ref{Opt1}), for all possibilities of $\cal{F}$.
However, we propose two structured class of strategies, Wait-to-Encode and Wait-to-Decode, and show that they are both able to achieve optimal delay scaling as channel rate $c$ approaches the long-term average conditional entropy $\mathbb{E}[H_{(t)}(X|Y)]$. Also in both classes of strategies, we provide ways to jointly encode and decode multiple (can be single if needed) blocks of source symbols together. In the rest of this section, our proposed strategies are presented in details.

\subsection{Wait-to-Encode Strategies}

A strategy, $\pi_{WE}$, is called Wait-to-Encode if it accumulates blocks of symbols at the encoder and these blocks are jointly encoded simultaneously. With Wait-to-Encode, a source block is encoded only in a single message and no longer kept at the encoder after encoding. At the end of each block $t$, the encoder makes a decision about generating message $M_t$, which is based on whether a condition associated with strategy $\pi_{WE}$ is satisfied or not. This condition is parameterized with $\textit{C}(\pi_{WE})$. Any time a block, $\mathbf{X}_t$, is deferred for future encoding, the associated message, $M_t$, at that instant is blank. Node $Y$ decodes $M_t$ immediately after receiving it, thus $W_{D}^{(\pi_{WE})}(t) \equiv 0$. Denoting the set of Wait-to-Encode strategies with $\Pi_{WE}$, we can summarize the general procedure as follows:

\vspace{0.15in}
\hrule
\vspace{0.1in}
\begin{algorithm}[Wait-to-Encode Strategy]\hfill

\noindent\textbf{Observation:}\\
At the beginning of block $t$, suppose $K-1$ blocks of source symbols have been accumulated thus far, for $K=1,2,\ldots$, waiting to be encoded. By the end of block $t$, node $X$ and $Y$ observes $\mathbf{x}_t$ and $\mathbf{y}_t$ respectively and the marginals $F_{(t)}(x)$ and $F_{(t)}(y)$ are interchanged between nodes.

\noindent\textbf{Encoding:}\\
Node $X$ generates message $M_t$ as follows:
                \begin{description}
                  \item[If \textbf{$\textit{C}(\pi_{WE})$} holds for the accumulated set of blocks]
                  \item[] Blank message $M_t$ is generated, i.e. $R_t = 0$;
                  \item[Else]
                  \item[] Message $M_t$ is generated via SW encoding at rate $R_t = R_X(F_X^K,F_Y^K,\epsilon)$,
               \end{description}
where $F_X^K$ and $F_Y^K$ represent the sequence of marginals for the observations of node $X$ and node $Y$, respectively and the encoding rate function $R_X(F_X^K,F_Y^K,\epsilon)$ is chosen as:
\begin{align}
R_X(F_X^K,F_Y^K,\epsilon)= &\min ~R_X      \label{JE}\\
                    \nonumber   &s.t. \hspace{-0.1in} \sum\limits_{F^K \in {\cal{F'}}^K}{P_F(F^K|F_X^K,F_Y^K) \le \epsilon},
\end{align}
where ${\scriptstyle{
{\cal{F'}}^K\triangleq \left\{F^K \in {\cal{F}}^K:\sum\limits_{\tau=t-K+1}^{t}H_{(\tau)}(X|Y)\ge R_X \right\}}}$. SW encoding is a random coding strategy, in which $2^{\sum_{\tau=t-K+1}^{t}H_{(\tau)}(\mathbf{X})}$ possible typical node $X$ observations are mapped into a binning structure with $2^{nR_X(F_X^K,F_Y^K,\epsilon)}$ bins. Then, given the vector of observations, node $X$ finds the bin number of the associated vector in the binning structure and uses it as message $M_t$.
\end{algorithm}
\vspace{0.1in}
\hrule
\vspace{0.15in}

\begin{theorem}
Given $\mathbf{x}^K$, $\mathbf{y}^K$, $F_X^K$, $F_Y^K$ and the outage probability constraint $\epsilon > 0$, the minimum achievable joint encoding rate for $M_t$ is the solution of the constrained optimization problem (\ref{JE}).
\end{theorem}

\begin{IEEEproof}
The detailed proof can be found in the \textit{Appendix}. 
To prove this theorem, we utilized random coding and typicality decoding ideas.
\end{IEEEproof}

\noindent {\bf Possible choices for Condition $\textit{C}(\pi_{WE})$:} It is clear that the main differentiator between different Wait-to-Encode schemes is the choice of Condition $\textit{C}(\pi_{WE})$. For instance, the two extreme cases for this class of algorithms are the ones in which $\textit{C}(\pi_{WE})$ is chosen such that (i) $K=\infty$ and (ii) $K=1$, which correspond to Illustrative Scenario (2) and (3), respectively. In general, Condition $\textit{C}(\pi_{WE})$ dictates the achievable point in the tradeoff between $\mathbb{E}[W_E^{(\pi_{WE})}(t)]$ and $\mathbb{E}[W_C^{(\pi_{WE})}(t)]$. If we try to keep one of them small, then the other will increase. With this observation, we propose the following Wait-to-Encode strategy $\pi_{WE}^*$, in which the condition $\textit{C}(\pi_{WE}^*)$ is chosen to be:
\begin{equation}
\label{eq:wait_to_encode_condition}
\frac{1}{K} R_X(F_X^K,F_Y^K,\epsilon) > c.
\end{equation}
Though $\pi_{WE}^*$ is not necessarily optimal with respect to Problem~(\ref{Opt1}), we prove in Section~\ref{PA} that it achieves optimal delay scaling as channel rate $c$ approaches the expected entropy rate, $\mathbb{E}[H_{(t)}(X|Y)]$, of the observation at node $X$.

\subsection{Wait-to-Decode Strategy}
In a Wait-to-Decode strategy, $\pi_{WD}$, the encoder generates message $M_t$ with encoding rate $R_t = c$, at the end of each time block $t$. For some blocks, this rate will be sufficient to decode $\mathbf{X}_t$ at the desired probability of outage. For those blocks that acquire higher encoding rates, block $\mathbf{X}_t$ is not dropped at the encoder, but jointly encoded with the subsequent blocks of source observations. Once the encoder decides that sufficient information is accumulated at the decoder so that the group of blocks encoded can be decoded with the desired probability of outage, the blocks at the encoder are removed and the next block is encoded by itself, to start the process afresh. With Wait-to-Decode strategy, the main delay is experienced at the decoder, since the messages are accumulated there. Once the delay is calculated at the decoder, the other components of delay can be written as $W_E^{(\pi_{WD})}(t) \equiv 1$ and $W_C^{(\pi_{WD})}(t) \equiv \frac{n\cdot R_t}{n\cdot c} = 1$ time block (due to the encoding rate $R_t = c$).

Motivated by the sequential binning strategy proposed in~\cite{Cheng}, we propose the following Wait-to-Decode strategy:

\vspace{0.15in}
\hrule
\vspace{0.1in}
\begin{algorithm}[Wait-to-Decode Strategy]\hfill

\noindent\textbf{Observation:}\\
At the beginning of block $t$, suppose $K-1$ blocks of source symbols and messages have been respectively accumulated at node $X$ and $Y$ thus far, for $K=1,2,\ldots$, waiting to be decoded. By the end of block $t$, node $X$ and $Y$ observes $\mathbf{x}_t$ and $\mathbf{y}_t$ respectively and the marginals $F_{(t)}(x)$ and $F_{(t)}(y)$ are interchanged between nodes.

\noindent\textbf{Encoding:}\\
Node $X$ generates message $M_t$ for jointly encoding $\mathbf{x}^K$ via sequential SW encoding at rate $R_t = c$. Sequential SW encoding is also a random coding strategy, in which $2^{\sum_{\tau=t-K+1}^{t}H_{(\tau)}(\mathbf{X})}$ possible typical node $X$ observations are mapped into a binning structure as with $2^{n\cdot c}$ bins. Then, given the vector of observations, node $X$ finds the bin number of the associated vector in the binning structure and uses it as message $M_t$. After accumulating the bin number sequence, i.e. message sequence $(M_{t-K+1},...M_t)$, this sequence of bin numbers is sequentially connected and treated as the `bin number' in traditional SW coding.

\noindent\textbf{Decoding:}\\
After receiving message $M_t$, node $Y$ decodes the accumulated messages $(M_{t-K+1},...M_t)$ as follows:
                \begin{description}
                  \item[If \textbf{$\frac{1}{K}R_X(F_X^K,F_Y^K,\epsilon) \leq c$} holds ]
                  \item[] Jointly decode all the accumulated messages $(M_{t-K+1},...M_t)$, remove $\mathbf{x}^K$ from the encoder and start afresh;
                  \item[Else]
                  \item[] All the messages $(M_{t-K+1},...M_t)$ keep being accumulated at node $Y$, waiting for the subsequent message(s).
               \end{description}

\end{algorithm}
\vspace{0.1in}
\hrule
\vspace{0.15in}
Notice that, we described only one Wait-to-Decode strategy, without any control condition. The main reason is that, here, we use a constant encoding rate (as opposed to the {\em variable-rate encoding} in Wait-to-Encode class of strategies) and at the same time node $Y$ decides to decode the accumulated messages as soon as the probability of outage goes below the desired threshold. We show in Section~\ref{PA} that our Wait-to-Decode strategy also achieves optimal delay scaling as channel rate $c$ goes down to $\mathbb{E}[H_{(t)}(X|Y)]$.

\subsection{Comparison of Wait-to-Encode and Wait-to-Decode}

\vspace{-0.05in}

Both proposed strategies have their own advantages and disadvantages. Under our Wait-to-Encode strategy $\pi_{WE}$, we carefully control the encoding rate, which is variable. We choose Condition $\textit{C}(\pi_{WE}^*)$ in a way to minimize the encoding delay by transmitting the group of blocks, as soon as the encoding rate goes below the channel rate. The optimal design of $\textit{C}(\pi_{WE})$ is complicated due to the strong coupling between the encoder and the channel, where the instant arrival rate of the channel, i.e. $R_t$, depends on our control decisions.

On the other hand, under the Wait-to-Decode strategy $\pi_{WD}$, we only have messages accumulating at the decoder. Due to the lack of control in our scheme, the end-to-end expected delay can be fairly large for certain blocks (despite being order optimal). For example, once a (very-long) block, demanding large SW encoding rate is observed, i.e., the required encoding rate is way above $c$, the decoder needs to wait for many subsequent blocks for things to smooth out in the long term. This is identical to the so called {\em slow truck effect} in First-in-First-out type queue scheduling.

%% file: Bounds.tex
\section{Performance Bounds}\label{PA}

In this section, we derive upper and lower bounds for the end-to-end delay under both Wait-to-Encode strategy $\pi_{WE}$ and Wait-to-Decode strategy $\pi_{WD}$. 

While our results are {\em general} for all possible values of the parameters, one of our main focus will be on the case in which the channel transmission rate $c$ is larger than, but close to $\mathbb{E}[H_{(t)}(X|Y)]$. To formalize this, we define parameter $\eta$ with $0<\eta<1$ such that 
\begin{eqnarray}
c = \frac{1}{(1-\eta)} \cdot \mathbb{E}[H_{(t)}(X|Y)]. 
\end{eqnarray}
The channel rate $c$ can be varied (but remains constant once given) by adjusting the value of the parameter $\eta$. The cases in which $\eta$ is close to $0$, i.e., $c\approx \mathbb{E}[H_{(t)}(X|Y)]$ are referred to as the {\em heavy traffic} regime. Heavy-traffic regime is particularly interesting in cases when we want to fully utilize the available channel, while achieving a low finite end-to-end delay. The results in this section reveal that both strategies achieve the order-optimal delay performance with respect to Problem (\ref{Opt1}) in heavy-traffic regime, i.e., $\eta\downarrow 0$. At the end of the section, we illustrate the bounds via a simple numerical example.


Recall that the set of possible joint cdfs $\cal{F}$ $ = \{ F_{11},\dots F_{1l_1}; F_{21},\dots F_{2l_2}; \dots; F_{m1},\dots F_{ml_m} \}$. Note that if $l_i=1, \forall 1\leq i\leq m$, then the marginal distribution information is equivalent to know the joint cdf information, for which the delay analysis becomes trivial. In fact, as we will discuss later, the delay scaling changes in the case of the availability of joint cdf, pointing out a {\em phase transition} phenomenon. The main focus in this section is in the case where there exists an $m^*$ with $1\leq m^{*}<m$ such that $l_i=1$ for any $i\leq m^*$ and $l_i>1$ for $m^{*}+1<i\leq m$. To simplify notations, we make the following definitions: 
\begin{align*}
\phi_{ij} &\triangleq P_F(F_{ij}), ~ \phi_i \triangleq  \sum_{j=1}^{l_i}\phi_{ij}, \\
H_{\max_i}(X|Y) &\triangleq \max_{1\leq j \leq l_i}H_{ij}(X|Y), 
\end{align*}
where $\phi_i$ and $H_{\max_i}$ denote the sum probability and the maximum conditional entropy of all joint cdfs in group $i$.

Next, we define some quantities that we use in expressing the performance. For all $(i,j)$ such that $1\leq i \leq m,~1 \leq j \leq l_i$, let
\begin{eqnarray*}
\sigma_{H}^2 &\triangleq& \sum\limits_{i=1}^{m}\sum\limits_{j=1}^{l_i}\phi_{ij}\left(H_{ij}(X|Y)-\mathbb{E}[H_{(t)}(X|Y)]\right)^2, \\
E_{H_i} &\triangleq& \sum\limits_{j=1}^{l_i}\frac{\phi_{ij}}{\phi_i}H_{ij}(X|Y), ~ c_i \triangleq \frac{1}{1-\eta} \cdot E_{H_i}, \\
\sigma_{H_i}^2 &\triangleq& \sum\limits_{j=1}^{l_i}\frac{\phi_{ij}}{\phi_i}\left(H_{ij}(X|Y)- E_{H_i} \right)^2.
\end{eqnarray*}
Note that, $\sigma_{H}^2$ is defined as the variance of the conditional entropies caused by variations in the joint cdff. $E_{H_i}$ and $\sigma_{H_i}^2$ denote the normalized expectation and variance of the conditional entropies associated with the joint cdfs in group $i$. Also, let us succinctly denote the objective function of optimization problem (\ref{Opt1}) with $\mathbb{E}[\overline{W}^{(\pi)}(t)]$, i.e., \vspace{-0.05in}
$$
\mathbb{E}[\overline{W}^{(\pi)}(t)] \triangleq \limsup \limits_{T \to \infty} \frac{1}{T} \sum\limits_{t=1}^{T}\mathbb{E}[{W_E^{(\pi)}(t)+W_C^{(\pi)}(t)+W_D^{(\pi)}(t)}] .
$$

\vspace{-0.1in}
\subsection{Delay Upper Bounds}
\vspace{-0.05in}

In this section, we provide upper bounds on the end-to-end delay achieved by Wait-to-Encode strategy $\pi_{WE}^*$ and Wait-to-Decode strategy $\pi_{WD}$ to be valid for all values of $\eta$.

\begin{theorem}\label{theoremUB}
For a given set of possible joint cdfs, $\cal{F}$,
expected average end-to-end delay $\mathbb{E}[\overline{W}^{(\pi_{WE}^*)}(t)]$ and $\mathbb{E}[\overline{W}^{(\pi_{WD})}(t)]$ can be upper bounded as follows:
\begin{eqnarray}
\mathbb{E}[\overline{W}^{(\pi_{WE}^*)}(t)] &\leq& \frac{3\cdot \gamma}{2} \cdot \frac{1}{\eta^2}+\frac{1}{2}, \label{UBWE}\\
\mathbb{E}[\overline{W}^{(\pi_{WD})}(t)] &\leq& \frac{\gamma}{2} \cdot \frac{1}{\eta^2}+\frac{3}{2}, \label{UBWD}
\end{eqnarray}
where 
\begin{align*}
\gamma &= \frac{-2 \ln{\epsilon}\cdot \sigma_H^2}{\mathbb{E}[H_{(t)}(X|Y)]^2}\left[(1-\eta)^2+\frac{M_H \mathbb{E}[H_{(t)}(X|Y)]}{3\sigma_H^2}\cdot (\eta - \eta^2)\right],\\
 M_H &= H_{\max}(X|Y)-\mathbb{E}[H_{(t)}(X|Y)].
 \end{align*}

\end{theorem}

\begin{IEEEproof}
Both strategies achieve their upper performance bounds in the worst case scenario when the encoder and the decoder do not use the marginal cdf of source symbols (or the marginal cdfs are useless, e.g. $m=1$, $l_1>1$). In such case, the joint encoding rate function $R_X(F_X^K, F_Y^K, \epsilon)$ no longer depends on $F_X^K$ or $F_Y^K$, but the value of $K$. Hence, $R_X(F_X^K, F_Y^K, \epsilon)$ can be simply written as $R_X(K,\epsilon)$. Based on this fact, we define constant $K_c$ as: \vspace{-0.1in}
\begin{eqnarray}
K_{c} \triangleq & \min & K     \label{KCCX}\\
\nonumber & \text{s.t.} & \mathbb{P} \left\{ \sum_{\tau=t-K+1}^{t} H_{(\tau)}(X|Y)>K\cdot c \right\} \leq \epsilon .
\end{eqnarray}
Since $\{ H_{(\tau)}(X|Y),\ \tau\geq 0\}$ is an i.i.d. process with mean $\mathbb{E}[H_{t}(X|Y)]$, which is less than c, there always exists a solution to (7) for all $\eta \in (0,1)$. Thus, it is clear that $R_X(K_c,\epsilon) \leq c$.
Thus, in Wait-to-Encode strategy $\pi_{WE}^*$, the encoder will always accumulate $K_c$ blocks to jointly encode; while in Wait-to-Decode strategy $\pi_{WD}$, the decoder will always accumulate $K_c$ blocks to jointly decode (because of the lack of side information from the observation of the marginals). Since every $K_c$ blocks form a cycle which will repeat over and over again in both strategies, the expected average end-to-end delay under the worst case scenario, i.e. $\mathbb{E}[\overline{W}_{WC}^{(\pi_{WE}^*)}(t)]$ and $\mathbb{E}[\overline{W}_{WC}^{(\pi_{WD})}(t)]$, can be easily derived as follows: 
\begin{align}
\mathbb{E}[\overline{W}_{WC}^{(\pi_{WE}^*)}(t)] &= \frac{1}{K_c}\sum\limits_{\tau=t-K_c+1}^{t}\left[W_E(\tau)+W_C(\tau)+W_D(\tau)\right] \nonumber \\
&= \frac{3K_c}{2}+\frac{1}{2}, \label{WCWE}
\end{align}
since $W_E(\tau)=t-\tau+1$, $W_C(\tau) \equiv \frac{c\cdot K_c}{c}$, and $W_D(\tau) \equiv 0$;

\vspace{-0.2in}
\begin{align}
\mathbb{E}[\overline{W}_{WC}^{(\pi_{WD})}(t)]&=\frac{1}{K_c}\sum\limits_{\tau=t-K_c+1}^{t}\left[W_E(\tau)+W_C(\tau)+W_D(\tau)\right]\nonumber \\
&=\frac{K_c}{2}+\frac{3}{2},    \label{WCWD}
\end{align}
since $W_E(\tau)=W_C(\tau) \equiv 1$, and $W_D(\tau)=t-\tau$.

In general, $K_c$ is difficult to be exactly evaluated. Yet, we can make use of Chernoff bound to approximately calculate $K_c$, denoted by $\widetilde{K}_c$. From Theorem 2.11 in \cite{Chernoff}, we have
\begin{multline*}
\mathbb{P} \left\{ \sum_{\tau=t-K+1}^{t} H_{(\tau)}(X|Y)>K\cdot c \right\} \\ \leq \exp\left\{-\frac{K\eta^2c^2}{2(\sigma_H^2+M_H\cdot c\cdot \eta/3)}\right\}.
\end{multline*}

Therefore, if we define $\widetilde{K}_c$ as:
\begin{eqnarray*}
\widetilde{K}_{c} = & \min & K     \\
\nonumber & \text{s.t.} & \exp\left\{-\frac{K\eta^2c^2}{2(\sigma_H^2+M_H\cdot c\cdot \eta/3)}\right\} \leq \epsilon ,
\end{eqnarray*}
then we have
\begin{eqnarray}
\widetilde{K}_c=\frac{-2 \ln{\epsilon}\cdot \sigma_H^2}{c^2\cdot \eta^2}\cdot \left(1+\frac{M_H\cdot c}{3\sigma_H^2}\cdot \eta\right).\label{TKCCX}
\end{eqnarray}
%
%
%
%

By substituting (\ref{TKCCX}) into (\ref{WCWE}) and (\ref{WCWD}), replacing $K_c$ and $c$ with $\widetilde{K}_c$ and $\frac{\mathbb{E}[H_{(t)}(X|Y)]}{1-\eta}$ respectively, we complete the proof.
\end{IEEEproof}

Note that $\gamma$ scales as $O(1)$ as $\eta$ approaches $0$. Theorem \ref{theoremUB} indicates that without any marginal distribution information, the expected end-to-end delay under both WE and WD strategies scales as $O(1/\eta^2)$. 
Next, we derive lower bounds for the same delay and show that the lower bounds also have the same scaling law. This indicates that our algorithms are order optimal.

\subsection{Delay Lower Bounds}
Next, we evaluate lower bounds on the end-to-end expected delay achieved for all $\pi_{WE} \in \Pi_{WE}$ and $\pi_{WD}$.
\begin{theorem}\label{theoremLB}
Given the set of possible joint cdfs $\cal{F}$,
expected delays $\mathbb{E}[\overline{W}^{(\pi_{WE})}(t)]$ and $\mathbb{E}[\overline{W}^{(\pi_{WD})}(t)]$ can be lower bounded as: \vspace{-0.15in}
\begin{multline}
\mathbb{E}[\overline{W}^{(\pi_{WE})}(t)] \geq \max \limits_{m^*+1\leq i^* \leq m} \left\{
(1-\eta)\sum\limits_{i \neq i^*}\frac{\phi_i E_{H_i}}{\mathbb{E}[H_{(t)}(X|Y)]} \right. \\
+\left. \phi_{i^*}\left(\frac{4 \gamma_{i^*}\cdot E_{H_{i^*}}}{27\cdot \mathbb{E}[H_{(t)}(X|Y)](1+\beta_{i^*})^2}\cdot \frac{1}{\eta^2}+\frac{1}{6}\right)
\right\} \label{LBWE}
\end{multline}

\vspace{-0.15in}

\begin{multline}
\mathbb{E}[\overline{W}^{(\pi_{WD})}(t)] \geq \max \limits_{m^*+1\leq i^* \leq m} \left\{ (1-\eta)\sum \limits_{i \neq i^*} \frac{\phi_i E_{H_i}}{\mathbb{E}[H_{(t)}(X|Y)]} \right. \\
\left. + \phi_{i^*}\left(\frac{2 \gamma_{i^*}\cdot E_{H_{i^*}}}{27\cdot \mathbb{E}[H_{(t)}(X|Y)](1+\beta_{i^*})^2}\cdot \frac{1}{\eta^2}+\frac{1}{2}\right)
\right\} , \label{LBWD}
\end{multline}
where
$\gamma_{i^*} = \frac{-2 \ln{\epsilon}\cdot \sigma_{H_{i^*}}^2}{E_{H_{i^*}}^2}\cdot \left[(1-\eta)^2+\frac{M_{H_{i^*}}\cdot E_{H_{i^*}}}{3\sigma_{H_{i^*}}^2}\cdot (\eta-\eta^2)\right]$, $M_{H_{i^*}} = H_{\max_{i^*}}(X|Y)-E_{H_{i^*}}$, and $\beta_{i*} = \frac{\sum\limits_{i \neq i^*}\phi_i E_{H_i}}{\phi_{i^*}E_{H_{i^*}}}$.
\end{theorem}

\begin{IEEEproof}
We only provide the detailed derivation of the lower bound for $\pi_{WE}$, since the lower bound of $\pi_{WD}$ can be derived by following identical steps.

First, we propose a genie-aided strategy whose delay performance lower bounded all Wait-to-Encode strategies. Fix any $i^*$ such that $m^*+1 \leq i^* \leq m$. We define this genie-aided scenario corresponding to $i^*$ in which the instantaneous joint cdf $F_{(t)}$ is provided to both of the nodes by a genie, if $F_{(t)} \notin \{F_{i^*1},...F_{i^*l_{i^*}}\}$, i.e. $F_{(t)}$ does not belong to the $i^*$th group of possible joint cdfs. Hence, all strategies can achieve as good performances, if no better, by exploiting more information about the joint source statistics. Consequently, the expected delay achieved by $\pi_{WE}$ in the genie-aided scenario, denoted by $\mathbb{E}[\overline{W}_{GC_{i^*}}^{(\pi_{WE})}(t)]$, can serve as the lower bound of the delay performance achieved by all $\pi_{WE}\in\Pi_{WE}$, i.e., 
\begin{eqnarray}
\mathbb{E}[\overline{W}^{(\pi_{WE})}(t)] &\geq& \mathbb{E}[\overline{W}_{GC_{i^*}}^{(\pi_{WE})}(t)]. \label{GCWE}
\end{eqnarray}

Next, we further provide the lower bound on $\mathbb{E}[\overline{W}_{GC_{i^*}}^{(\pi_{WE})}(t)]$:
\begin{multline}
\label{GCLBLB}
\mathbb{E}[\overline{W}_{GC_{i^*}}^{(\pi_{WE})}(t)] \geq (1-\eta)\sum\limits_{i \neq i^*}\frac{\phi_i c_i}{c} \\ +
\phi_{i^*} \sum\limits_{K=1}^{\infty}\alpha_K^{(\pi_{WE})} \left(\frac{K+1}{2}+\frac{ R_X(F_{X_{i^*}}^K,F_{Y_{i^*}}^K,\epsilon)}{c}\right) ,
\end{multline}
where $\alpha_K^{(\pi_{WE})}$ denotes the empirical probability such that K blocks, with joint pdfs belong to the $i^*$ group, are jointly encoded under strategy $\pi_{WE}$. This inequality holds, since we take the following procedures to further reduce the end-to-end delay:
\emph{1)} For the blocks with joint cdf $F_{ij}$ ($i \neq i^*$), decode them block by block with encoding rate $H_{ij}(X|Y)$ and only take the transmission delay into account , i.e. $\frac{H_{ij}(X|Y)}{c}$;  \emph{2)} Assume that the channel is always idle and ready to serve whenever a message arrives.

For $R_X(K,\epsilon) = R_X(F_{X_{i^*}}^K,F_{Y_{i^*}}^K,\epsilon)$, we further derive the lower bound of the second term in (\ref{GCLBLB}) as follows:
\begin{eqnarray}
&\nonumber &  \sum\limits_{K=1}^{\infty}\alpha_K^{(\pi_{WE})} \cdot \left(\frac{K+1}{2}+\frac{R_X(K,\epsilon)}{c}\right)  \\
& \stackrel{(a)}{=}&  \sum\limits_{K=1}^{K'-1}\alpha_K^{(\pi_{WE})} \cdot \left(\frac{K+1}{2}+\frac{R_X(K,\epsilon)}{c}\right)+\nonumber \\
&\nonumber& \sum\limits_{K=K'}^{\infty}\alpha_K^{(\pi_{WE})} \cdot \left(\frac{K+1}{2}+\frac{R_X(K,\epsilon)}{c}\right)   \\
& \stackrel{(b)}{\geq}& \sum\limits_{K=1}^{K'-1}\alpha_K^{(\pi_{WE})} \cdot \frac{\left(1+r(1+\beta_{i^*})\eta\right)c_{i^*}}{c} + \nonumber \\ &\nonumber& \sum\limits_{K=K'}^{\infty}\alpha_K^{(\pi_{WE})} \cdot \left(\frac{K'+1}{2}+\frac{K' \cdot (1-\eta)c_{i^*}}{c}\right)  \\
& \stackrel{(c)}{=}& \frac{\alpha\cdot c_{i^*}}{c} \cdot \left(1+r(1+\beta_{i^*})\eta\right) + \nonumber \\ &\nonumber& \frac{(1-\alpha)\cdot c_{i^*}}{c} \cdot \left(\frac{(K'+1)\cdot c}{2c_{i^*}} + K'(1-\eta)\right)  \\
& \stackrel{(d)}{\geq}& \frac{\cdot c_{i^*}}{c\cdot(1+r)} \cdot \left(1+r(1+\beta_{i^*})\eta\right) + \nonumber \\ &\nonumber& \frac{r\cdot c_{i^*}}{c\cdot (1+r)} \cdot \left(\frac{(K'+1)\cdot c}{2c_{i^*}} + K'(1-\eta)\right)  \\
& \stackrel{(e)}{\geq}& \frac{4c_{i^*}}{27c}\cdot \gamma_{i^*} \cdot \frac{1}{(1+\beta_{i^*})^2\cdot \eta^2}+\frac{1}{6} , \label{GCLBLBLB}
\end{eqnarray}
where step ($a$) holds for $K'=K_{\left(1+r(1+\beta_{i^*})\eta\right)c_{i^*}}$, in which constant $K_{\left(1+r(1+\beta_{i^*})\eta\right)c_{i^*}}$ is defined in the same way as in (\ref{KCCX}) by replacing $c$ with $\left(1+r(1+\beta_{i^*})\eta\right)c_{i^*}$, for some $r>0$ and $F_{(\tau)} \in \{F_{i^*1},...F_{i^*l_{i^*}}\}$; step ($b$) follows from the fact that $\frac{1}{K}R_X(K,\epsilon)\geq \left(1+r(1+\beta_{i^*})\eta\right)c_{i^*}$ for $K<K'$, and the fact that $\frac{1}{K}R_X(K,\epsilon)\geq (1-\eta)c_{i^*}$ (the equality can only be achieved when $K \to \infty$); step ($c$) is true for taking $\alpha=\sum\limits_{K=1}^{K'-1}\alpha_K^{(\pi_{WE})}$; step ($d$) holds when $\alpha$ achieves its maximum, $\frac{1}{1+r}$, under the average encoding rate constraint: $\sum\limits_{K=1}^{\infty}\alpha_K^{(\pi_{WE})}\cdot \frac{R_X(K,\epsilon)}{K} \leq \left(1+(1+\beta_{i^*})\eta\right)c_{i^*}$; step ($e$) follows by setting $K'=\widetilde{K}'$, and further bounding $\widetilde{K}'$ from below by $\widetilde{K}'\geq \gamma_{i^*}\cdot\frac{1}{(1+r)^2(1+\beta_{i^*})^2\eta^2}$ via similar steps we have used in the proof of \textit{Theorem}~\ref{theoremUB} with $r=\frac{1}{2}$.

By substituting (\ref{GCLBLBLB}) into (\ref{GCLBLB}), taking the maximum over all $i^*$, combining with inequality \eqref{GCWE}, and replacing $c$, $c_i$ and $c_{i^*}$ with $\frac{\mathbb{E}[H_{(t)}(X|Y)]}{1-\eta}$, $\frac{E_{H_i}}{1-\eta}$ and $\frac{E_{H_{i^*}}}{1-\eta}$ respectively, we have the desired result. 

\end{IEEEproof}

Note that $\gamma_{i^*}$ scales as $O(1)$ when $\eta$ approaches $0$. Therefore, Theorem \ref{theoremLB} reveals that the end-to-end delay scales as $\Omega(1/\eta^2)$ under both WE and WD strategies. 

\subsection{Delay Scaling with Known Joint Distribution}
From Theorem \ref{theoremUB} and Theorem \ref{theoremLB}, we showed that both WE and WD strategies achieve order-optimal delay performance that scales as $O(1/\eta^2)$ in the heavy-traffic regime, i.e., $\eta\downarrow 0$. Also, one interesting observation is that the marginal distribution information does not necessarily improve the delay scaling. This is in contrast to the case where the perfect knowledge of the joint cdf is available. 

\begin{figure}
\vspace{-0.0in}
\begin{center}
{\includegraphics[height=0.8in]{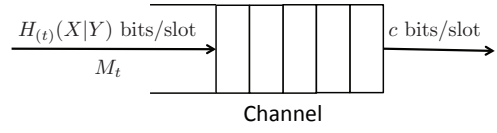}} \vspace{-0.0in}
\caption{Source symbols are enqueued as messages $M_t$, at the input of the channel for transmission at fixed rate $c$ bits/slot.}
\label{Traditional_Queueing_Model}
\end{center}
\vspace{-0in}
\end{figure}

When the joint cdf is known, which refers to Illustrative Scenario (1), traditional SW encoding and decoding can be applied on a block-by-block basis. Hence, each source block experiences unit time block delay and zero delay at the encoder and decoder, respectively, i.e.:
\begin{eqnarray}
W_{E}^{(\pi)}(t) \equiv 1, ~ W_{D}^{(\pi)}(t) \equiv 0. \label{Delay_TRQ}
\end{eqnarray}
Also, source block $\mathbf{X}_t$ is encoded as $M_t$ at rate $H_{(t)}(X|Y)$, and then transmitted over the channel. Thus, we have a queue at the input of the channel with random arrivals $H_{(t)}(X|Y)$ and constant service rate $c$ (bits/slot), as presented in Fig. \ref{Traditional_Queueing_Model}. It can be proved that the expected delay experienced in the channel, $\mathbb{E}[W_{C}^{(\pi)}(t)]$, scales with $O(1/\eta)$ (via a direct application of \cite[Lemma 4]{erysri12}, which analyzes the delay scaling in a single FIFO queue) in the heavy-traffic regime, i.e. $\eta \downarrow 0$. Together with (\ref{Delay_TRQ}), we can conclude that the expected end-to-end delay $\mathbb{E}[W^{\pi}(t)]$ scales with $(1/\eta)$. This points out a {\bf phase transition phenomenon}: Even a minor degradation in the knowledge of joint statistics at the encoder -from perfect knowledge to a slightly imperfect knowledge- leads to a different scaling regime in the expected delay.


\subsection{Numerical Evaluation}
\begin{figure}[h]
  \center
     \includegraphics[scale = 0.4]{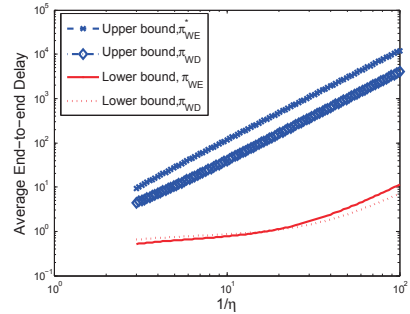}
     \vspace{-0.1in}
    \caption{Delay bounds for Wait-to-Encode and Wait-to-Decode schemes: both x-axis and y-axis are in logarithmic scale.}
\vspace{-0.1in}
    \label{NumericalEvaluation}
 \end{figure}

To illustrate how the performance bounds vary with $\eta$, we study a simple example as follows: (i) $\cal{F}$ $=\{F_{11},F_{21},F_{22},F_{31},F_{32},F_{33}\}$, $\epsilon=0.01$; (ii) $\phi_{11}=0.1$, $\phi_{21}=\phi_{22}=0.2$, $\phi_{31}=0.12$, $\phi_{32}=\phi_{33}=0.19$; (iii) $H_{ij}(X|Y)=i+j$. Thus the source entropy varies between $2-6$ bits/symbol. In  Fig~\ref{NumericalEvaluation}, we plot the upper and lower bounds on the expected delay for the observations taken at node $X$ to be replicated at node $Y$, measured in number of blocks ($n$ slots), as a function of the heavy-traffic parameter inverse $1/\eta$.

Examining the plots, one can see that both upper and lower bounds scale as $O(1/\eta^2)$ as $\eta \to 0$, which reflects the order optimality of our proposed strategies. Also, for the above set of parameters, the bounds slightly favor Wait-to-Decode strategy, but note that, this is not a typical trend and one can come up with another set of parameters for which the opposite holds.

%% file: Experiment.tex
\section{Experimental Evaluation}\label{EE}

To illustrate the performance of average end-to-end delay and compare the proposed encoding and decoding strategies, we set up an experiment which emulates a possible application of our problem in real-time streaming networks. As shown in Fig.~\ref{EF} and discussed in Section~\ref{Intro}, we set two cameras at different locations at still to record people walking by in a busy street. The distance between the cameras is $20$ meters and the directions they face are at a $90^{\circ}$ angle to one another.

After synchronization, we end up with two correlated uncompressed video frame sequences. For the video signals, each frame has a size $1024\times 1024$, where each pixel is represented with $3$ bytes (associated with RGB-index), hence each frame is a $3$ MB sample. The frame rate is $60$~frame/sec, i.e. the streaming rate of the uncompressed video is $180$~Mbps. We regard each frame as one video symbol (a very large one) and pile up every $n=180$ symbols to form blocks $\{\mathbf{X_t}\},\ \{\mathbf{Y_t}\}$ in our model, i.e., a time slot is $1/60$ secs long and a block is $3$ secs long.
The empirically generated time-varying joint (marginal) cdf sequences are represented by $\{F_{(t)}(x,y)\}$, $\{F_{(t)}(x)\}$, and $\{F_{(t)}(y)\}$.

To form the set ${\cal F}$, we use a pilot shot over a certain duration. Due to the huge size of the symbols and the unpredictable environment, developing the exact representation of set ${\cal F}$ is not possible. Thus, we quantize the set of all possible distributions: Firstly, using the pilot sequence, we calculate the relative entropy of each block $t$ with respect to block $1$, i.e. $D\left(F_{(t)}(x,y)||F_{(1)}(x,y)\right)$, which gives us the sequence of  relative entropies $\{D_{(t)}(x,y)\}$. Next, we quantize the values of $\{D_{(t)}(x,y)\}$ into $128$ intervals and treat each quantization level as a single joint distribution, one which is randomly picked from each interval. Hence, we end up with $128$ different quantized joint cdfs $\{F_{ij}(x,y)\}$ in order to form the set $\cal{F}$. Finally, we repeat an identical quantization process for the marginal cdf sequences $\{F_{ij}(x)\}$ and $\{F_{ij}(y)\}$, but we set the number of quantization intervals to $8$ this time. Thus, we have $64$ different combinations of marginal cdf pairs. Recall that $\cal{F}$ $ = \{ F_{11},\dots F_{1l_1}; F_{21},\dots F_{2l_2}; \dots; F_{m1},\dots F_{ml_m} \}$, hence $m=64$ after the above quantization process. The empirical PMF $P_F(F_{ij})$ can be calculated with respect to each $F_{ij} \in \cal{F}$, using the pilot shot. Hence, we use an {\bf extremely coarse quantizer} in the representation of the sources. One of our main objectives is to show that, even with such coarse quantization, it is possible to achieve a substantial compression ratio (e.g., reduction in the rate of data transmission from node $X$ to node $Y$) for $\mathbf{X}_t$ to be replicated at node $Y$. After the pilot shot, we first quantize the observed new frames as described above. After quantization, the conditional entropy rate we observed from the combined pilot shots turned out to be $76.5$~Mbps, which we take as the basic limit for the minimum rate node $X$ needs to transmit for lossless recovery ($0$-outage) at node $Y$. Note that, once we obtain the video traces based on real-world data and construct set ${\cal F}$, we use simulations to evaluate the performance of the system. We ran these real-world data-driven simulations for $10^6$ blocks of video symbols.

\noindent
\begin{figure*}[htp!]
\vspace{-0.1in}
\centerline{
\subfloat[Delay performance of proposed strategies]
{\includegraphics[width=2.45in]{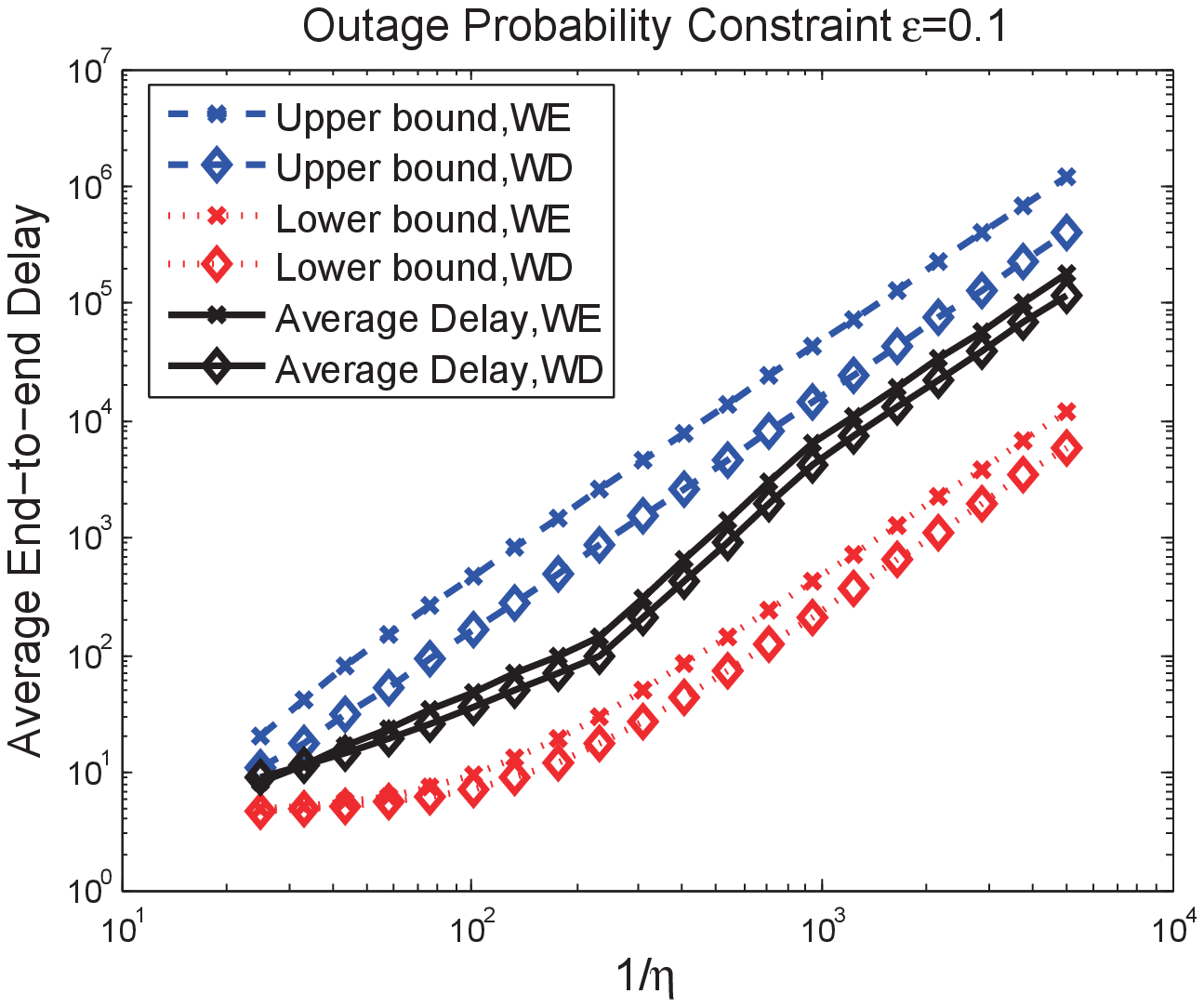}
\label{HeavyTraffic}}
\subfloat[Encoding rate under strategy $\pi_{WE}^*$]
{\includegraphics[width=2.45in]{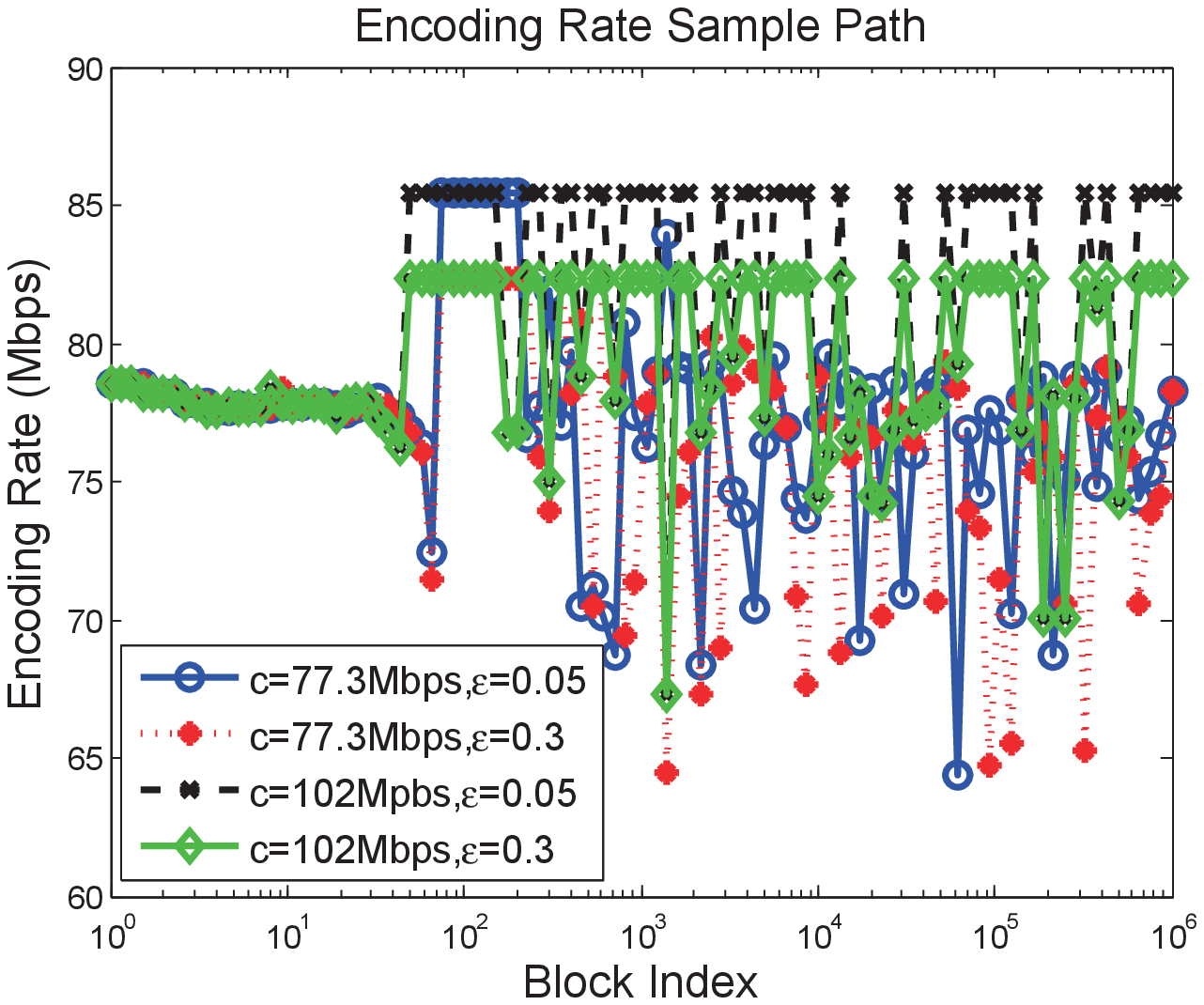}
\label{RateSample}}
\subfloat[End-to-end block delay under $\pi_{WE}^*$ and $\pi_{WD}$]
{\includegraphics[width=2.45in]{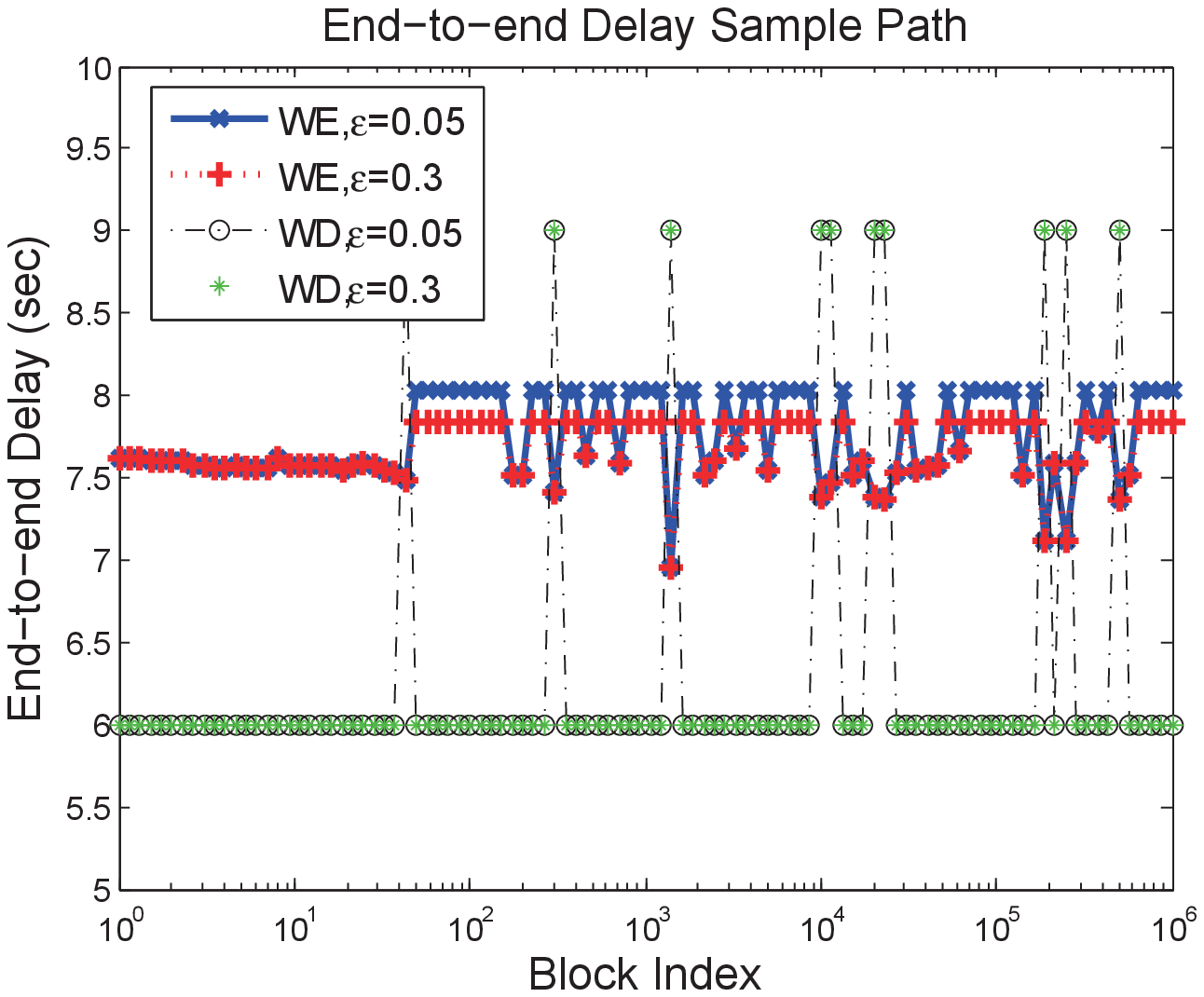}
\label{DelaySample}}
}
\caption{Experimental Results. With a delay of a few blocks, the compression ratio of around $50$\% is achievable for lossless replication subject to a probability of outage of $0.05$.}
\vspace{-0.2in}
\end{figure*}

In our first evaluation, we focus on the heavy-traffic scenario. Transmission rate is fixed at $c=\frac{1}{1-\eta}\times 76.5$~Mbps, for a given heavy-traffic parameter $\eta$. The end-to-end delay experienced by each block was stored and then used to calculate the average delay over all $10^6$ blocks. Average delay is plotted vs. $\eta^{-1}$ in Fig.~\ref{HeavyTraffic} for Wait-to-Encode strategy $\pi_{WE}^*$, Wait-to-Decode strategy $\pi_{WD}$, along with their associated upper/lower bounds we derived in Section~\ref{PA}. In the plot, `WE' and `WD'stand for `Wait-to-Encode' and `Wait-to-Decode'. 

As can be observed in Fig.~\ref{HeavyTraffic}, the expected delay of all proposed strategies scale as $O(\frac{1}{\eta^2})$ as $\eta \to 0$. Note that, even though $\pi_{WD}$ achieves better performance than $\pi_{WE}^*$ in our experiment, this is not necessarily a common trend for all possible Wait-to-Encode strategies.  Also, it is worth mentioning that the bounds become tighter, as the traffic load gets lower.


Next, we plot the encoding rate chosen by our Wait-to-Encode strategy, $\pi_{WE}^*$, as a function of time, for various values of $\eta$ and outage probability constraint $\epsilon$. Note that, in Wait-to-Decode strategy, $\pi_{WD}$, the encoding rate is always  identical to $c$, which we choose to be $\frac{1}{1-\eta}\times 76.5$ Mbps in these plots. In Fig.~\ref{RateSample}, we illustrate the encoding rate, measured regularly across blocks. In particular, we take one sample per $10^4$ blocks through the entire trace. The channel rate (thus the encoding rate for $\pi_{WD}$) is $c=102$ and $77.3$ Mbps for $\eta=0.75$ and $0.99$, respectively. One can see in Fig~\ref{RateSample}, if the traffic is light (i.e., the average conditional entropy of the source is much smaller than the channel rate), $\pi_{WE}^*$ chooses the encoding rate high, since the fixed channel transmission rate is sufficiently large to choose encoding rate more aggressively to reduce delay. Also, as expected, a smaller value of $\epsilon$ requires higher encoding rate to keep the outage constraint to be met. Finally, note that the long-term average encoding rate with $\pi_{WE}^*$ will always be smaller than that with $\pi_{WD}$, but the ratio of the average encoding rates will be no less than $1-\eta$.

Another interesting observation from this plot is the following.  Recall that the rate of the uncoded video is $180$ Mbps; thus, with the available channel rates, it is not possible for the uncoded video to be transmitted from node $X$ to node $Y$. However, even with the highly coarse quantization that we use for the source observations, we reduced the average encoding rate (i.e., the rate at which node $X$ transmits to node $Y$) to between $75$-$85$ Mbps at an outage probability of $0.05$, which corresponds to a compression ratio of approximately $50$\% at a reasonably low block delay, which will be analyzed next.

Finally, we focus on the delay for two different values of $\epsilon$ at $\eta=0.25$, which corresponds to a $75$\% utilization. We plot the sample path for the end-to-end block delay in Fig.~\ref{DelaySample} for strategies $\pi_{WE}^*$ and $\pi_{WD}$. In these experiments, $c=102$ Mbps. With only a delay of $6$-$9$ secs, we achieve an outage probability of $0.05$. This delay achieved, combined with the $50$\% compression ratio demonstrates the efficacy of our system. We believe much higher compression ratios could be possible with finer quantizers and higher correlations between sources.
Finally, note that the end-to-end delay becomes constant as the transmission rate increases (low utilization), leading to Illustrative scenario (3), depicted in Section~\ref{PS}. In that case, block-by-block encoding becomes the best scheme.

\noindent {\bf Discussion:} There is another alternative for our system design that we did not consider here, as it pertains to video compression. The observed video signal at node $X$ can be compressed via standard video compression techniques (non-distributed), independently of the observation of node $Y$. The basic limit for the rate after that process is $\mathbb{E}[H_{(t)}(X)]$, while one can achieve a much smaller rate $\mathbb{E}[H_{(t)}(X|Y)]$ (i.e., a much higher compression ratio) with the distributed approach. 
One important question that we are planning to answer as a part of future work is the comparison of our schemes with standard video compression. We will study systematic ways to develop quantizers (possibly more complex than we have here) that are simple, yet provide significant gains over the best available non-distributed video compression techniques. 

%% file: Conclusion.tex
\section{Conclusions}\label{CD}
We studied the lossless distributed source coding problem in which there exists a pair of nodes, observing a random source with time-varying statistics, unknown to the nodes before the session starts. We formulated the problem as that of minimization of end-to-end delay for the observations of one of the nodes to be replicated at the other node, subject to a certain desired outage probability. Even though, it is not possible to come up with well-structured solutions to the basic problem, due to the generality of the set of distributions, we developed two different classes of strategies, Wait-to-Encode and Wait-to-Decode that are {\em provably} order optimal in the heavy traffic limit. After analytically deriving general bounds for the expected delay achieved by our schemes, we further experimentally demonstrate the efficacy of the schemes, using a setup involving two cameras, obtaining videos of a common scene at different angles. We showed that, even with a very low-complexity quantizer, a compression ratio of approximately $50$\% is achievable for lossless replication at the decoder, at an average delay of $5$-$10$ seconds.

%% file: Appendice.tex
\appendix
\section{Proof of \textit{Theorem} $1$}

\begin{IEEEproof}
Without loss of generality, in the proof, we replace `$t$' and `$t-K+1$' in the statement of the theorem with `$t+K-1$' and `$t$' respectively.

First, we define the typical sequences $A_\epsilon^{Kn}(X|F_X^K)$, $A_\epsilon^{Kn}(Y|F_Y^K)$ and joint typical sequence $A_\epsilon^{Kn}(X,Y|F^K)$ as following:
\begin{eqnarray}
A_\epsilon^{Kn}(X|F_X^K) &=& \left\{\mathbf{x}^K:|\overline{S'}_X^K-\overline{S}_X^K|\leq \epsilon \right\} \nonumber \\
A_\epsilon^{Kn}(Y|F_Y^K) &=& \left\{\mathbf{y}^K:|\overline{S'}_Y^K-\overline{S}_Y^K|\leq \epsilon \right\} \nonumber \\
A_\epsilon^{Kn}(X,Y|F^K) &=&\left\{ (\mathbf{x}^K, \mathbf{y}^K):
\begin{array}{c l}
                         |\overline{S'}_X^K-\overline{S}_X^K|\leq \epsilon  \nonumber \\
                         \nonumber |\overline{S'}_Y^K-\overline{S}_Y^K|\leq \epsilon \\
                         \nonumber |\overline{S'}_{XY}^K-\overline{S'}_{XY}^K|\leq \epsilon
\end{array}
                         \right\}
\end{eqnarray}
where
\begin{align*}
\overline{S'}_X^K=-\frac{1}{Kn}\sum\limits_{\tau=t}^{t+K-1}\log P_{(\tau)}(\mathbf{x}_{\tau}), \\ \overline{S}_X^K=\frac{1}{K}\sum\limits_{\tau=t}^{t+K-1}H_{(\tau)}(X), \\ \overline{S'}_Y^K=-\frac{1}{Kn}\sum\limits_{\tau=t}^{t+K-1}\log P_{(\tau)}(\mathbf{y}_{\tau}),\\ \overline{S}_Y^K=\frac{1}{K}\sum\limits_{\tau=t}^{t+K-1}H_{(\tau)}(Y),\\ \overline{S'}_{XY}^K=-\frac{1}{Kn}\sum\limits_{\tau=t}^{t+K-1}\log P_{(\tau)}(\mathbf{x}_{\tau},\mathbf{y}_{\tau}),\\ \overline{S}_{XY}^K=\frac{1}{K}\sum\limits_{\tau=t}^{t+K-1}H_{(\tau)}(X,Y).
\end{align*}

And the (joint) typical sequences have following properties:
\begin{enumerate}
  \item \begin{flalign*}
   2^{-n\big(S_X^K\big)^+} \leq \prod\limits_{\tau=t}^{t+K-1} P_{(\tau)}(\mathbf{x}_{\tau}) \leq 2^{-n\big(S_X^K\big)^-} \nonumber\\
   2^{-n\big(S_Y^K\big)^+} \leq \prod\limits_{\tau=t}^{t+K-1} P_{(\tau)}(\mathbf{y}_{\tau}) \leq 2^{-n\big(S_Y^K\big)^-} \nonumber\\
   2^{-n\big(S_{XY}^K\big)^+} \leq \prod\limits_{\tau=t}^{t+K-1} P_{(\tau)}(\mathbf{x}_{\tau},\mathbf{y}_{\tau}) \leq 2^{-n\big(S_{XY}^K\big)^-} \nonumber
  \end{flalign*}
  \item \begin{align}
  \mathbb{P}\left\{(\mathbf{x}^K,\mathbf{y}^K) \in A_\epsilon^{Kn}(X,Y|F^K)\right\} \xrightarrow{n \rightarrow \infty}1 \nonumber \\
  \mathbb{P}\left\{\mathbf{x}^K \in A_\epsilon^{Kn}(X|F_X^K)\right\} \xrightarrow{n \rightarrow \infty}1 \nonumber \\
  \mathbb{P}\left\{\mathbf{y}^K \in A_\epsilon^{Kn}(Y|F_Y^K)\right\} \xrightarrow{n \rightarrow \infty}1 \nonumber
  \end{align}
  \item \begin{align*}
  (1-\epsilon)\cdot 2^{n\big(S_X^K\big)^-} \leq |A_\epsilon^{Kn}(X|F_X^K)| \leq 2^{n\big(S_X^K\big)^+} \nonumber\\
  (1-\epsilon)\cdot 2^{n\big(S_Y^K\big)^-} \leq |A_\epsilon^{Kn}(Y|F_Y^K)| \leq 2^{n\big(S_Y^K\big)^+} \nonumber\\
  (1-\epsilon)\cdot 2^{n\big(S_{XY}^K\big)^-} \leq |A_\epsilon^{Kn}(X,Y|F^K)| \leq 2^{n\big(S_{XY}^K\big)^+} \nonumber
  \end{align*}
  \item $\forall ~\mathbf{y}^K \in A_\epsilon^{Kn}(Y|F_Y^K)$,
  \begin{eqnarray}
  |A_\epsilon^{Kn}(X|F^K,\mathbf{y}^K)| \leq 2^{n\big(S_{X|Y}^K\big)^+}\nonumber
  \end{eqnarray}
\end{enumerate}
where
\begin{align*}
S_X^K=\sum\limits_{\tau=t}^{t+K-1}H_{(\tau)}(X),\\
S_Y^K=\sum\limits_{\tau=t}^{t+K-1}H_{(\tau)}(Y),\\
S_{XY}^K=\sum\limits_{\tau=t}^{t+K-1}H_{(\tau)}(X,Y),\\
S_{X|Y}^K=\sum\limits_{\tau=t}^{t+K-1}H_{(\tau)}(X|Y),\\
(S)^-=S-\delta(\epsilon),~(S)^+=S+\delta(\epsilon).
\end{align*}
And $\delta(\epsilon)>0$ is a function of $\epsilon$ satisfying: $\delta(\epsilon) \xrightarrow{\epsilon \to 0} 0$. Property $1$ follows by the definition of (joint) typical sequences; Property $2$ can be proved by laws of large numbers(LLN); Property $3$ follows by the first two properties; Property $4$ can be achieved with the first three properties.

Second, we focus on the encoding and decoding process for Wait-to-Encode and Wait-to-Decode strategies:\\
\noindent {\bf Wait-to-Encode Strategy:} Codebooks are generated according to the number of blocks K, thus we have a sequence of codebooks: $\{C_K\}$. To generate each codebook $C_K$, all possible sequences $\mathbf{x}^K$ are uniformly distributed into $2^{nR_{t+K-1}}$ bins, where $R_{t+K-1}$ is the chosen joint encoding rate for $M_{t+K-1}$. These codebooks are shared between the encoder and decoder beforehand. Upon receiving the $X$ symbol sequence $\mathbf{x}^K$, the encoder checks whether this is a typical sequence. If so, the encoder finds out the bin number, denoted as $B(\mathbf{x}^K)$ and shortly as $B_K$, in which the sequence locates. Otherwise, set $B_K$ equal to 1. Then the encoder sends $B_K$ to the decoder as encoded message, i.e. $M_{t+K-1} = B_K$. Upon receiving the encoded message $B_K$, the decoder picks each sequence $\mathbf{\hat{x}}^K$ in the bin $B_K$ which satisfies $\mathbf{\hat{x}}^K \in A_\epsilon^{Kn}(X|F_X^K)$, and tests if there exists $\hat{F}^K$ such that the following two conditions hold: (1) $(\mathbf{\hat{x}}^K,\mathbf{y}^K) \in A_\epsilon^{Kn}(X,Y|\hat{F}^K)$; (2) $P_F(\hat{F}^K|F_X^K,F_Y^K)>0$. If there exits more than one sequence $\mathbf{\hat{x}}^K$ with its corresponding $\hat{F}^K$ satisfying the aforementioned two conditions, then select the sequence with highest value of $P_F(\hat{F}^K|F_X^K,F_Y^K)$ as the decoded message. And if there exists no such sequence, the decoder reports an failure of decoding.\\
\noindent {\bf Wait-to-Decode Strategy:} For $k=1,2,...K$, codebooks are generated according to the number $k$, thus we have a sequence of codebooks: $\{C^k\}$. To generate each codebook $C^k$, all possible sequences $\mathbf{x}^k$ are uniformly distributed into $2^{n\cdot c}$ bins, where $c$ is the chosen joint encoding rate for $M_{t+k-1}$, i.e. $R_{t+k-1}=c$. These codebooks are shared between the encoder and decoder beforehand. Upon receiving the $X$ symbol sequence $\mathbf{x}^k$, the encoder checks whether this is a typical sequence. If so, the encoder finds out the bin number, denoted as $B(\mathbf{x}^k)$ and shortly as $B^k$, in which the sequence $\mathbf{x}^k$ locates. Otherwise, set $B^K$ equal to 1. Then the encoder sends $B^k$ to the decoder as the encoded message, i.e.$M_{t+k-1}=B^k$. Upon receiving the encoded message sequence $\{B^k\}$, the decoder picks each sequence $\mathbf{\hat{x}}^K$, which has the same bin number sequence $\{B_k\}$ and satisfies $\mathbf{\hat{x}}^K \in A_\epsilon^{Kn}(X|F_X^K)$, and tests if there exists $\hat{F}^K$ such that the following two conditions hold: (1) $(\mathbf{\hat{x}}^K,\mathbf{y}^K) \in A_\epsilon^{Kn}(X,Y|\hat{F}^K)$; (2) $P_F(\hat{F}^K|F_X^K,F_Y^K)>0$. If there exits more than one sequence $\mathbf{\hat{x}}^K$ with its corresponding $\hat{F}^K$ satisfying the aforementioned two conditions, then select the sequence with highest value of $P_F(\hat{F}^K|F_X^K,F_Y^K)$ as the decoded message. And if there exists no such sequence, the decoder reports an failure of decoding.

Third, we finalize the proof with decoding error analysis for both strategies. There exists three decoding error events, denoted as $\varepsilon_1$, $\varepsilon_2$ and $\varepsilon_3$, which are defined as following:
\begin{eqnarray}
\varepsilon_1 &=& \left\{(\mathbf{x}^K,\mathbf{y}^K) \not \in A_\epsilon^{Kn}(X,Y|F^K) \right\} \nonumber \\
\varepsilon_2 &=& \left\{
\begin{array}{c l}\exists \mathbf{\hat{x}}^K \neq \mathbf{x}^K: \\ \hat{B}_K=B_K,(\mathbf{\hat{x}}^K,\mathbf{y}^K) \in A_\epsilon^{Kn}(X,Y|F^K)
\end{array}
 \right\} \nonumber \\
\varepsilon_3 &=& \left\{
\begin{array}{c l}
\exists \mathbf{\hat{x}}^K  \neq \mathbf{x}^K , \hat{F}^K \neq F^K: \\ B_K=\hat{B}_K,(\mathbf{\hat{x}}^K ,\mathbf{y}^K ) \in A_\epsilon^{Kn}(X,Y|\hat{F}^K)
\end{array}
\right\} \nonumber
\end{eqnarray}
where $\hat{B}_K=B(\mathbf{\hat{x}}^K)$, $F^K$ denotes the true joint cdf sequence. \\
\noindent {\bf Wait-to-Encode Strategy:} By asymptotic equipartition property(AEP), we have: $\mathbb{P}\{\varepsilon_1\} \to 0$ as $n \to \infty$.
\begin{eqnarray}
\mathbb{P}\{\varepsilon_2\} &\leq & \sum\limits_{(\mathbf{x}^K ,\mathbf{y}^K )}\prod\limits_{\tau=t}^{t+K-1} P_{(\tau)}(\mathbf{x}_{\tau},\mathbf{y}_{\tau})\cdot \nonumber \\
                            &\nonumber& \sum\limits_{\mathbf{\hat{x}}^K  \neq \mathbf{x}^K :\mathbf{\hat{x}}^K  \in A_\epsilon^{Kn}(X|F^K,\mathbf{y}^K)}\mathbb{P}\{\hat{B}_K=B_K\} \nonumber \\
            \nonumber       &\leq & \sum\limits_{(\mathbf{x}^K ,\mathbf{y}^K )}\prod\limits_{\tau=t}^{t+K-1} P_{(\tau)}(\mathbf{x}_{\tau},\mathbf{y}_{\tau})\cdot \frac{|A_\epsilon^{Kn}(X|F^K,\mathbf{y}^K)|}{2^{nR_{t+K-1}}} \\
            \nonumber       &\leq & \frac{1}{2^{nR_{t+K-1}}} \cdot 2^{n\big(S_{X|Y}^K + \delta(\epsilon) \big)} \\
            \nonumber       & = & 2^{-n\big(R_{t+K-1}-S_{X|Y}^K-\delta(\epsilon)\big)}
\end{eqnarray}
Thus, under the condition $R_{t+K-1}>S_{X|Y}^K+\delta(\epsilon)$, we have:  $\mathbb{P}\{\varepsilon_2\} \to 0$ as $n \to \infty$.
\begin{eqnarray}
\mathbb{P}\{\varepsilon_3\} &\leq & \sum\limits_{\hat{F}^K \neq F^K} P_F(\hat{F}^K|F_X^K,F_Y^K) \cdot \nonumber \\ &\nonumber& \sum\limits_{(\mathbf{x}^K,\mathbf{y}^K) \in A_\epsilon^{Kn}(X,Y|\hat{F}^K)} \prod\limits_{\tau=t}^{t+K-1} \hat{P}_{(\tau)}(\mathbf{x}_{\tau},\mathbf{y}_{\tau})  \cdot \\
                            &\nonumber & \sum\limits_{\mathbf{\hat{x}}^K \neq \mathbf{x}^K: \mathbf{\hat{x}}^K \in A_\epsilon^{Kn}(X|\hat{F}^K,\mathbf{y}^K)}\mathbb{P}\{\hat{B}_K=B_K\} \nonumber \\
   \nonumber                &\leq & \sum\limits_{\hat{F}^K \neq F^K} P_F(\hat{F}^K|F_X^K,F_Y^K) \cdot \\ &\nonumber& \sum\limits_{(\mathbf{x}^K,\mathbf{y}^K) \in A_\epsilon^{Kn}(X,Y|\hat{F}^K)} \prod\limits_{\tau=t}^{t+K-1} \hat{P}_{(\tau)}(\mathbf{x}_{\tau},\mathbf{y}_{\tau}) \cdot \\ &\nonumber& \frac{|A_\epsilon^{Kn}(X|\hat{F}^K,\mathbf{y}^K)|}{2^{nR_{t+K-1}}} \\
   \nonumber                &\leq & \frac{1}{2^{nR_{t+K-1}}} \cdot 2^{n\big(S_{X|Y}^K + \delta(\epsilon) \big)} \\
   \nonumber                & = & 2^{-n\big(R_{t+K-1}-S_{X|Y}^K -\delta(\epsilon)\big)}
\end{eqnarray}
Similarly, under the condition $R_{t+K-1}>S_{X|Y}^K+\delta(\epsilon)$, we have:  $\mathbb{P}\{\varepsilon_3\} \to 0$ as $n \to \infty$.
We define the outage probability $\mathbb{P}_{out}^K$ as: $\mathbb{P}_{out}^K \stackrel{\Delta}{=} \mathbb{P}\big\{\mathbf{\hat{X}}^K \neq \mathbf{X}^K \big\}$. Then we can derive the upper bound of $\mathbb{P}_{out}^K$ as following:
\begin{eqnarray}
\mathbb{P}_{out}^K &\leq & \mathbb{P}\{R_{t+K-1}>S_{X|Y}^K\} \cdot \big(\mathbb{P}\{\varepsilon_1\}+ \mathbb{P}\{\varepsilon_2\} +\mathbb{P}\{\varepsilon_3\}\big) \nonumber \\
                 &\nonumber& + \mathbb{P}\{R_{t+K-1} \leq S_{X|Y}^K\} \nonumber \\
  \nonumber      &\leq & (1-\epsilon)\cdot \big(\mathbb{P}\{\varepsilon_1\}+\mathbb{P}\{\varepsilon_2\}+\mathbb{P}\{\varepsilon_3\}\big) + \epsilon \\
  \nonumber      &\stackrel{n \to \infty}{=} & (1-\epsilon) \cdot 0 + \epsilon \\
  \nonumber      &=& \epsilon
\end{eqnarray}

Notice that $K$ can take any value of integers, and $\mathbb{P}_{out}^K$ for $\forall K$. Let $\alpha_K^{(\pi)}$ denote the empirical probability for a block to be jointly encoded with other $K-1$ blocks under strategy $\pi$, the overall outage probability $\mathbb{P}_{out}\stackrel{\Delta}{=} \mathbb{P}\big\{\mathbf{\hat{X}_t} \neq \mathbf{X_t} \big\}$, can be derived as following:
\begin{eqnarray*}
\mathbb{P}_{out} &\leq& \sum\limits_{K=1}^{\infty} \alpha_K^{(\pi)} \cdot \mathbb{P}_{out}^K \\
                 &\leq& \sum\limits_{K=1}^{\infty} \alpha_K^{(\pi)} \cdot \epsilon \\
                 &\leq& \epsilon
\end{eqnarray*}
Hence, with encoding rate $R_{t+K-1}$ for $M_{t+K-1}$, the outage probability constraint $\epsilon$ can be achieved.\\
\noindent {\bf Wait-to-Decode Strategy:} a similar decoding error analysis can be derived by replacing $B_K$ with $\{B^k\}$.

This completes the proof.
\end{IEEEproof}

%% file: Infocomm2015_JournalVersion.bbl
\begin{thebibliography}{10}

\bibitem{Cheng}
C.~Chang.
\newblock Streaming source coding with delay.
\newblock In {\em Thesis,EECS Department University of California, Berkeley},
  2007.

\bibitem{Fangzhou15}
F.~Chen, B.~Li, and C.~E. Koksal.
\newblock Low-delay distributed source coding for time-varying sources with
  unknown statistics.
\newblock Proceedings of the 34th IEEE International Conference on Computer
  Communications (INFOCOM) Conference, Hong Kong, China, Apr. 2015.

\bibitem{CK}
I.~Csiszar and J.~Korner.
\newblock Towards a general theory of source networks.
\newblock {\em IEEE Trans. on Inf. Th}, 26(2):155--165, Mar 1980.

\bibitem{Draper}
S.~Draper.
\newblock Universal incremental slepian-wolf coding.
\newblock In {\em Proc. 2nd Int. Conf. Quality Service Heterogeneous
  Wired/Wireless Networks}, 2004.

\bibitem{Elsa}
E.~Dupraz, A.~Roumy, and M.~Kieffer.
\newblock Source coding with side information at the decoder: Models with
  uncertainty, performance bounds, and practical coding schemes.
\newblock In {\em Information Theory and its Applications (ISITA)}, volume~40,
  pages 586--595, 2012.

\bibitem{erysri12}
A.~Eryilmaz and R.~Srikant.
\newblock Asymptotically tight steady-state queue length bounds implied by
  drift conditions.
\newblock {\em Queueing Systems}, 72:311--359, 2012.

\bibitem{Chernoff}
F.Chung and L.Lu.
\newblock Complex graphs and networks.
\newblock In {\em American Mathematical Society}, 2006.

\bibitem{Girod}
B.~Girod, A.Aaron, S.~Rane, and D.~Rebollo-Monedero.
\newblock Distributed video coding.
\newblock {\em Proceedings of the IEEE}, 93(1):71--83, Jan 2005.

\bibitem{Jaggi}
S.~Jaggi and M.~Effros.
\newblock Universal linked multiple access source codes.
\newblock In {\em Information Theory, Proceedings. IEEE International Symposium
  on}, 2002.

\bibitem{li2013heavy}
B.~Li, R.~Li, and A.~Eryilmaz.
\newblock Heavy-traffic-optimal scheduling with regular service guarantees in
  wireless networks.
\newblock In {\em ACM MOBIHOC, Bangalore, India}, July 2013.

\bibitem{Liveris02}
A.~Liveris, Z.~Xiong, and C.~Georghiades.
\newblock Compression of binary sources with side information at the decoder
  using ldpc codes.
\newblock {\em IEEE Commun. Lett.}, 6:440--442, 2002.

\bibitem{RM}
R.~Matsumoto.
\newblock Problems in application of ldpc codes to information in quantum key
  distribution protocols.
\newblock In {\em IEICE Technical Report.}, 2009.

\bibitem{Matsuta10}
T.~Matsuta, T.~Uyematsu, and R.~Matsumoto.
\newblock Universal slepian-wolf source codes using low-density parity-check
  matrices.
\newblock In {\em Proc. IEEE International Symposium on Information Theory},
  pages 186--190, 2010.

\bibitem{Reza}
R.~Parseh and F.~Lahouti.
\newblock Rate adaptation for slepian-wolf coding in presence of uncertain side
  information.
\newblock In {\em The Fifth International Conference on Sensor Technologies and
  Applications}, 2011.

\bibitem{Puri02}
R.~Puri and K.~Ramchandran.
\newblock Prism: a new robust video coding architecture based on distributed
  compression principles.
\newblock In {\em Proc. Allerton Conference on Communications Control and
  Computing}, pages 665--674, 2002.

\bibitem{SS}
S.~Shriram, B.~D., and B.~R.G.
\newblock Variable-rate universal slepian-wolf coding with feedback.
\newblock In {\em Signals, Systems and Computers, Conference Record of the
  Thirty-Ninth Asilomar Conference on}, pages 8--12, 2005.

\bibitem{SW}
D.~Slepian and J.~Wolf.
\newblock Noiseless coding of correlated information sources.
\newblock {\em IEEE Trans. on Inf. Th}, 19(4):471--480, Jul 1973.

\bibitem{Stankovic04}
V.~Stankovic, A.~D. Liveris, Z.~Xiong, and C.~N. Georghiades.
\newblock Design of slepian-wolf codes by channel code partitioning.
\newblock In {\em Proc. Data Compression Conference}, pages 302--311, 2004.

\bibitem{Stankovic06}
V.~Stankovic, A.~D. Liveris, Z.~Xiong, and C.~N. Georghiades.
\newblock On code design for the slepian-wolf problem and lossless
  multiterminal networks.
\newblock {\em IEEE Trans. Inf. Theory}, 52(4):1495--1507, 2006.

\bibitem{Xiong}
Z.~Xiong, A.~Liveris, and S.~Cheng.
\newblock Distributed source coding for sensor networks.
\newblock {\em Signal Processing Magazine, IEEE}, 21(5):80--94, Oct 2004.

\end{thebibliography}
